\documentclass[sigconf,natbib=true,anonymous=false]{acmart}

\usepackage{algorithmic}
\usepackage{multirow}
\usepackage{subcaption}
\usepackage{xspace}
\usepackage{colortbl}

\acmSubmissionID{641}

\newcommand{\eg}{\emph{e.g.,}\xspace}

\newcommand{\ignore}[1]{}

\newcommand{\model}{Fave\xspace}
\newcommand{\fullmodel}{\textbf{F}low-based \textbf{A}verage \textbf{V}elocity \textbf{E}stablishment\xspace}
\title{FAVE: Flow-based Average Velocity Establishment for Sequential Recommendation}
\definecolor{mygreen}{rgb}{0.0, 0.45, 0.0}
\definecolor{myred}{rgb}{0.7, 0.0, 0.0}
\definecolor{lightgray}{gray}{0.9}

\setcopyright{acmlicensed}
\copyrightyear{2026}
\acmYear{2026}
\acmConference[SIGIR '26]{Proceedings of the 49th International ACM SIGIR Conference on Research and Development in Information Retrieval}{July 20--24, 2026}{Melbourne, Australia}

\begin{document}


\author{Ke Shi\textsuperscript{1,2,\textdagger}, Yao Zhang\textsuperscript{1,\textdagger}, Feng Guo\textsuperscript{1}, Jinyuan Zhang\textsuperscript{1}, JunShuo Zhang\textsuperscript{1}, Shen Gao\textsuperscript{1,2,*}, Shuo Shang\textsuperscript{1,2,*}}
\affiliation{%
  \institution{\textsuperscript{1}University of Electronic Science and Technology of China\\
  \textsuperscript{2}State Key Laboratory of Internet Architecture}
  \country{}
}
\email{{keshi08.13lue15, yaoz2023, jinyuanegg, hanblingz026, jedi.shang}@gmail.com}
\email{zxyz0051@163.com, shengao@uestc.edu.cn}

\begin{abstract}
Generative recommendation has emerged as a transformative paradigm for capturing the dynamic evolution of user intents in sequential recommendation. 
While flow-based methods improve the efficiency of diffusion models, they remain hindered by the ``Noise-to-Data'' paradigm, which introduces two critical inefficiencies: prior mismatch, where generation starts from uninformative noise, forcing a lengthy recovery trajectory; 
and linear redundancy, where iterative solvers waste computation on modeling deterministic preference transitions.
To address these limitations, we propose a \fullmodel (\model) framework for one-step generation recommendation that learns a direct trajectory from an informative prior to the target distribution.
\model is structured via a progressive two-stage training strategy. 
In Stage 1, we establish a stable preference space through dual-end semantic alignment, applying constraints at both the source (user history) and target (next item) to prevent representation collapse. 
In Stage 2, we directly resolve the efficiency bottlenecks by introducing a semantic anchor prior, which initializes the flow with a masked embedding from the user's interaction history, providing an informative starting point. 
Then we learn a global average velocity, consolidating the multi-step trajectory into a single displacement vector, and enforce trajectory straightness via a JVP-based consistency constraint to ensure one-step generation.
Extensive experiments on three benchmarks demonstrate that \model not only achieves state-of-the-art recommendation performance but also delivers an order-of-magnitude improvement in inference efficiency, making it practical for latency-sensitive scenarios\footnote{Code is available at \url{https://github.com/Blue130/Fave}}.
\end{abstract}



\maketitle

\noindent\textsuperscript{\textdagger}Equal contribution. \textsuperscript{*}Corresponding authors.

\section{Introduction}

The core objective of sequential recommendation (SR) is to accurately model the dynamic evolution of user intent, reflected in user interaction sequences~\cite{17, 19, 29}. 
To achieve this, generative recommendation has emerged as a promising paradigm~\cite{27, 28}. 
By learning a direct mapping from historical behaviors to future preference distributions, it offers superior robustness compared to traditional discriminative approaches~\cite{94, 74}. 
For example, diffusion models~\cite{95, 93, 91} employ non-autoregressive generation to capture global dependencies and achieve high-fidelity preference modeling. 
However, their inference efficiency falls short of the strict real-time requirements in recommendation tasks.
The main reason lies in the iterative ``noise-to-data'' generation process, which requires numerous denoising steps to reconstruct a user profile from random noise~\cite{79, 78}. 
The resulting computational overhead presents a significant barrier in latency-sensitive applications~\cite{8, 26}.

\begin{figure}
    \centering
    \includegraphics[width=1\linewidth]{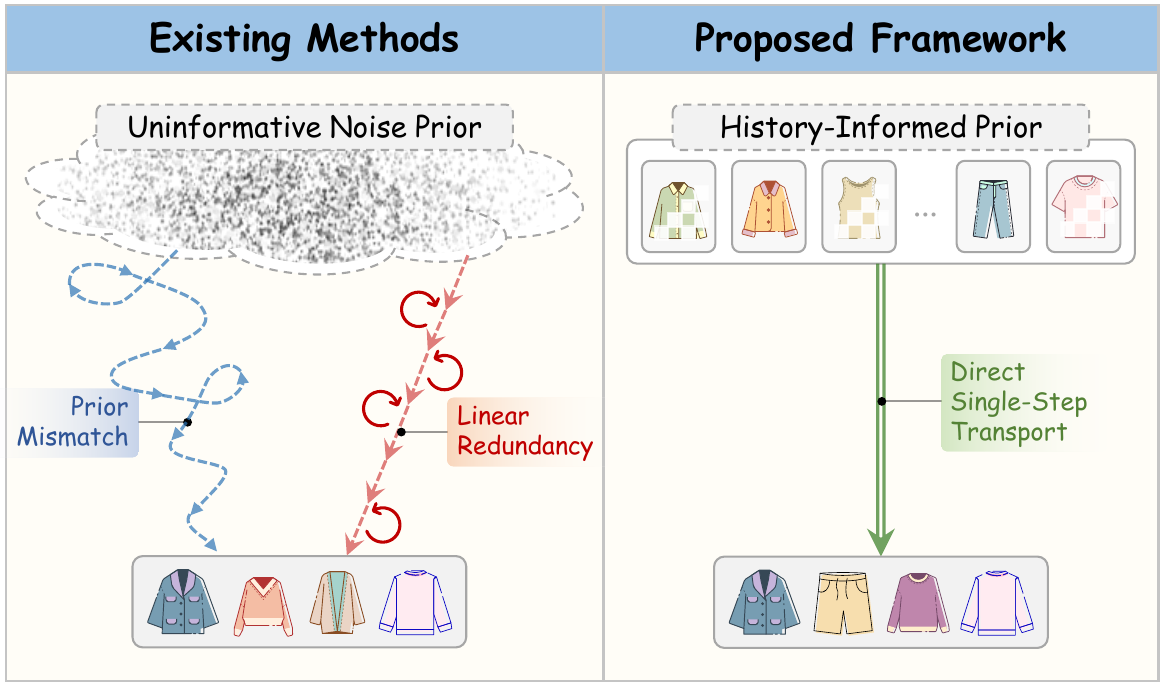}
    \caption{Comparison of generative recommendation paradigms. (Left) Existing ``Noise-to-Data'' generative methods suffer from prior mismatch and linear redundancy. (Right) Our proposed \model employs direct trajectory transport from a semantic anchor prior.
    }
    \label{fig:motivation}
\end{figure}

To mitigate this computational overhead, recent research has adopted the flow matching framework~\cite{10, 12, 24}. 
Unlike the stochastic denoising process of diffusion models~\cite{25, 98}, flow matching formulates the generation path as a deterministic trajectory governed by Ordinary Differential Equations (ODEs). 
By establishing a direct, linear interpolation between the source and target distributions, it theoretically constructs a significantly smoother trajectory, contrasting with the complex curvature inherent in diffusion-based generation~\cite{96}.
For instance, FMRec~\cite{10} reduces the need for constant directional correction during inference, thereby suppressing cumulative errors arising from trajectory curvature. This inherent linearity enables more efficient numerical integration, allowing the model to take larger, more confident steps toward the target, thus substantially reducing computational cost compared to diffusion models.
However, existing flow-based recommendation methods still exhibit notable inefficiencies.

The observed inefficiency originates from a fundamental misalignment between the conventional ``Noise-to-Data'' paradigm in flow matching and the intrinsic requirements of recommendation tasks. 
This misalignment gives rise to two primary bottlenecks:
\textbf{1. Prior Mismatch.} 
Current flow-based methods initialize the generation process from an uninformative noise distribution~\cite{10}. 
In contrast, the starting point should be a well-defined representation of user interest. 
Forcing the model to recover a user profile from noise necessitates traversing an excessively long trajectory during inference, consuming substantial computational resources. 
\textbf{2. Linear Redundancy.} 
In recommendation, flow matching converges toward a fixed target distribution~\cite{92, 84} by utilizing an ODE solver. 
Once the ODE solver determines the update direction, the trajectory's evolution direction remains invariant throughout inference. 
Consequently, each subsequent iterative step merely repeats the same deterministic transformation, leading to redundant computations without meaningful progress. 
Figure~\ref{fig:motivation} provides an intuitive illustration of these two challenges. 
The blue winding path exemplifies prior mismatch, showing the extended trajectory required to reconstruct user semantics from an uninformative start. 
The red dashed loops depict linear redundancy, where iterative solvers waste computation on repetitive updates along a fixed direction.

In this paper, we propose \model, a novel one-step generative recommendation framework that leverages \fullmodel to construct a direct trajectory from an informative prior to the target user preference distribution. 
To achieve this, \model employs a two-stage training strategy that decouples foundational representation learning from trajectory optimization. 
The first stage is basic manifold construction which utilizes flow matching to learn instantaneous velocity fields from a Gaussian prior. 
This phase focuses on stabilizing training and constructing a high-quality latent preference manifold.
The second stage is single-step consolidation which directly targets the identified inefficiencies. 
To overcome prior mismatch, we introduce a semantic anchor prior, which replaces the uninformative Gaussian noise with an initial state derived from the user's masked interaction history. 
This anchors the starting point near the target in the semantic space to shorten the trajectory.
To eliminate linear redundancy, we shift from instantaneous velocities to a global average velocity establishment. 
This velocity is regularized by a Jacobian-vector-product (JVP) based straightness constraint, which enforces consistency in both direction and evolution speed throughout the generation process. 
Thus, we can obtain a single-step displacement that directly transports the user's behavior pattern to the target item distribution.
Extensive experiments on three widely-used benchmarks demonstrate that \model achieves state-of-the-art recommendation performance while delivering an order-of-magnitude improvement in inference efficiency compared to strong baselines.

\noindent We summarize our contributions as follows:

\noindent $\bullet$ We propose \model, a one-step flow matching generative recommendation framework which constructs a direct trajectory, resolving the efficiency bottleneck.

\noindent $\bullet$ We introduce a semantic anchor prior to replace the uninformative Gaussian noise, which mitigates the inefficiency of uninformative initialization.

\noindent $\bullet$ We introduce a single-step consolidation mechanism via an average velocity field and impose a JVP-based straightness constraint to guarantee consistency in both speed and direction.

\noindent $\bullet$ Extensive experiments on three public datasets demonstrate \model achieves state-of-the-art recommendation performance and order-of-magnitude efficiency in inference.

\section{Related Work}
\paragraph{Sequential Recommendation}
Sequential recommendation (SR) aims to predict future user interactions based on historical behavior sequences~\cite{20, 21}. Traditional solutions predominantly adopt a discriminative approach. Early deep learning methods, such as GRU4Rec~\cite{2}, utilize RNN to model temporal dynamics, while Caser~\cite{3} employs CNN to capture local sequential patterns as image features. More recently, self-attention mechanisms have emerged as the dominant paradigm; SASRec~\cite{4} utilizes a unidirectional transformer to model long-range dependencies, while BERT4Rec~\cite{5} employs a bidirectional encoder trained via a masked item prediction objective to capture deeper context. To address the inherent uncertainty in user behaviors, STOSA~\cite{6} further represents items as stochastic Gaussian distributions using Wasserstein self-attention.


\paragraph{Flow Matching}
Flow matching (FM) has emerged as a compelling generative paradigm, employing continuous normalizing flows to define deterministic probability paths via Ordinary Differential Equations (ODEs)~\cite{1, 13, 83, 75}. Unlike diffusion models~\cite{7, 8, 9, 100}, FM learns a time-dependent velocity field to guide the generation process. In sequential recommendation, FlowRec~\cite{12} applies this framework to explicitly model user preference trajectories via a learned velocity field. Similarly, FMRec~\cite{10} adopts straight flow trajectories with a deterministic Euler sampler to minimize error accumulation from noise perturbations.

\section{Preliminary}
\paragraph{Generative Sequential Recommendation}
Let $\mathcal{U}=\{u_1, u_2, \dots, u_{|\mathcal{U}|}\}$ and $\mathcal{I}=\{i_1, i_2, \dots, i_{|\mathcal{I}|}\}$ represent the sets of users and items. The binary interaction matrix is defined as $\mathbf{A} \in \{0, 1\}^{|\mathcal{U}| \times |\mathcal{I}|}$, where $\mathbf{A}_{u,i}=1$ indicates observed interactions. For each user $u \in \mathcal{U}$, the historical behavior is recorded as a sequence $s_u = [i_1, i_2, \dots, i_l]$, where $i_l \in \mathcal{I}$ denotes the $l$-th item interacted with by user $u$. The objective is to predict the next interaction $i_{l+1}$ based on the history $s_u$. Distinct from traditional methods, generative sequential recommendation introduces a novel paradigm centered on probability distributions. Specifically, the model aims to learn a parameterized non-linear mapping $f_\theta$ that transforms a tractable prior distribution $p_{\text{prior}}$ directly into the target distribution $p_{\text{data}}$ which encapsulates user preferences:
\begin{equation}
p_{\text{data}} = f_\theta(p_{\text{prior}}).
\end{equation}

\paragraph{Flow Matching}
Flow matching formulates the problem as learning a time-dependent velocity field $v_\theta(\boldsymbol{x}_t, t)$. This field defines a probabilistic trajectory that transforms a prior distribution into a target distribution. 

Let $\boldsymbol{x}_0 \sim p_{\text{prior}}$ and $\boldsymbol{x}_1 \sim p_{\text{data}}$ denote samples drawn from the prior and data distributions, respectively~\cite{18}. The data state at time $t$ follows a linear interpolation trajectory,
\begin{equation}
\label{eq:define_inter}
\boldsymbol{x}_t = (1 - t)\boldsymbol{x}_0 + t \boldsymbol{x}_1.
\end{equation}
Under the linear interpolation trajectory, the conditional instantaneous velocity connecting a specific pair $(\boldsymbol{x}_0, \boldsymbol{x}_1)$ can be obtained by differentiation,
\begin{equation}
u_t(\boldsymbol{x}_t \mid \boldsymbol{x}_0, \boldsymbol{x}_1) = \frac{d \boldsymbol{x}_t}{d t} = \boldsymbol{x}_1 - \boldsymbol{x}_0.
\end{equation}

Since trajectories from different starting points may overlap at interpolated state $\boldsymbol{x}_t$, the direction of movement becomes ambiguous. Flow matching resolves this by learning a global velocity field that represents the average direction of all training pairs passing through this point:
\begin{equation}
\bar{v}(\boldsymbol{x}_t, t)
\triangleq \mathbb{E}_{p_t(\boldsymbol{x}_1, \boldsymbol{x}_0 \mid \boldsymbol{x}_t)}[u_t].
\end{equation}
To capture this velocity field, we employ a neural network parameterized by $\theta$, optimizing the following objective:
\begin{equation}
\mathcal{L}_{\mathrm{FM}}(\theta)
= \mathbb{E}_{t, p_t(\boldsymbol{x}_t)} \left\| v_\theta(\boldsymbol{x}_t, t) - \bar{v}(\boldsymbol{x}_t, t) \right\|^2.
\end{equation}
However, directly computing this target is intractable because the exact composition of trajectories passing through $\boldsymbol{x}_t$ is unknown during the training process. To circumvent this, following \cite{1}, we optimize the conditional flow matching loss:
\begin{equation}
\begin{aligned}
\label{eq:loss_cfm}
\mathcal{L}_{CFM}(\theta)
&= \mathbb{E}_{t,\, \boldsymbol{x}_0,\, \boldsymbol{x}_1}
\left\| v_\theta(\boldsymbol{x}_t, t) - u_t(\boldsymbol{x}_t \mid \boldsymbol{x}_0, \boldsymbol{x}_1) \right\|^2 \\
&= \mathbb{E}_{t,\, \boldsymbol{x}_0,\, \boldsymbol{x}_1}
\left\| v_\theta(\boldsymbol{x}_t, t) - (\boldsymbol{x}_1 - \boldsymbol{x}_0) \right\|^2 .
\end{aligned}
\end{equation}

\begin{figure*}
    \centering
    \includegraphics[width=1\textwidth]{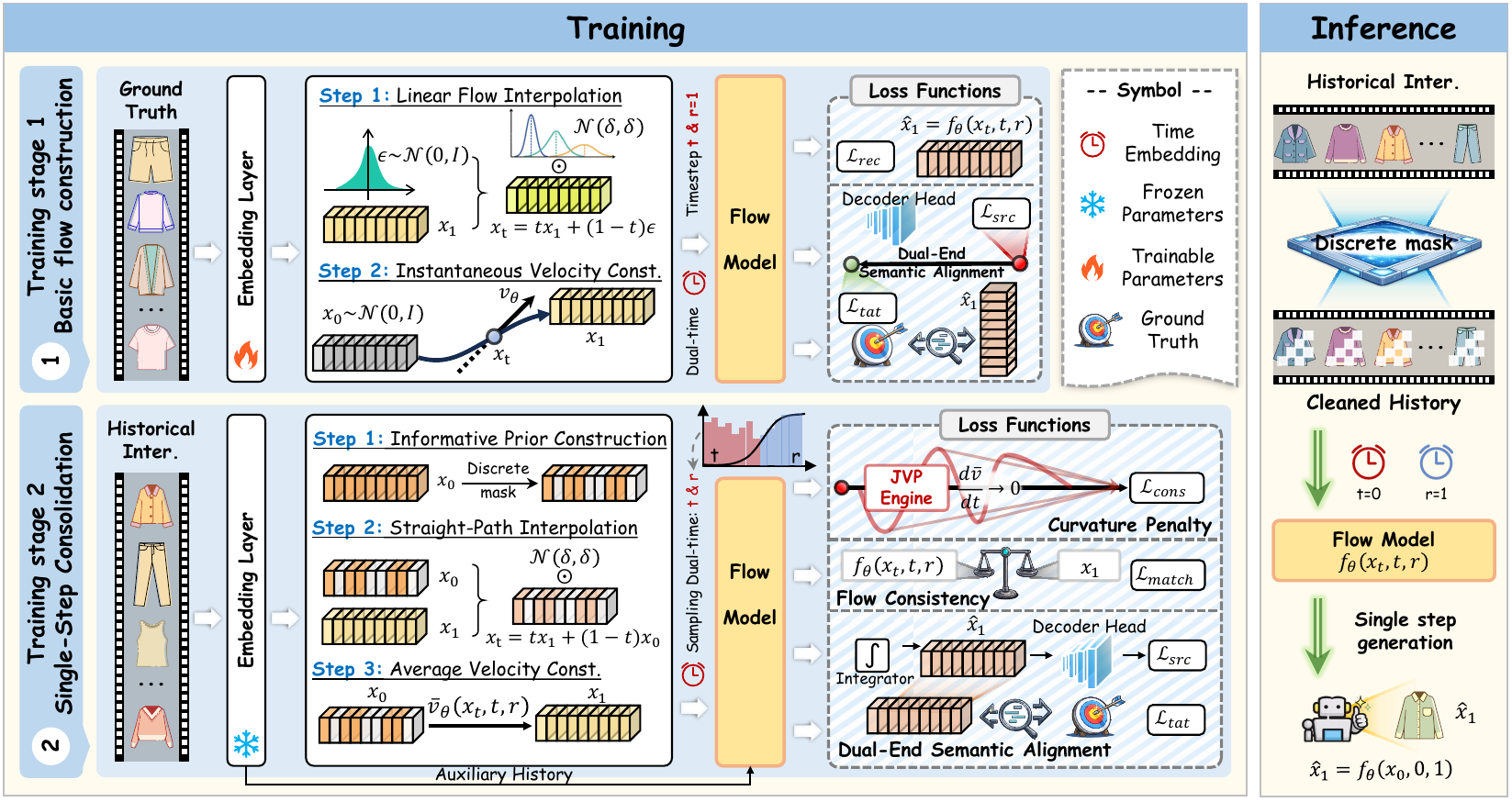}
    \caption{The overall framework of \model which adopts a two-stage training strategy. It first constructs a basic manifold by learning the instantaneous velocity field from Gaussian noise. The second stage incorporates a \textbf{semantic anchor prior} and \textbf{average velocity modeling} to address the prior mismatch and linear redundancy issue.
}
    \label{fig:framework}
\end{figure*}

\section{Methodology}
\subsection{Overall Framework}
To simultaneously address the prior mismatch and linear redundancy issues, we propose \model. 
As illustrated in Figure~\ref{fig:framework}, our framework adopts a progressive two-stage training strategy, consistently governed by a dual-end alignment mechanism to ensure the stability of the semantic space. 
The first stage involves building a basic manifold, learning the fundamental probability trajectory by studying the instantaneous velocity field originating from Gaussian noise.
The second stage utilizes a semantic anchor prior to minimize trajectory distance and models average velocity, thereby resolving the bottlenecks of prior mismatch and linear redundancy.

\subsection{Training Stage 1: Basic Manifold Construction}

To establish a robust generative foundation, this stage formulates the recommendation task as a trajectory generation problem on a continuous latent manifold.
Specifically, we implement this modeling with two main strategies:
First, to capture the precise dynamics of interest evolution, we introduce a dual time mechanism that estimates instantaneous velocity. This enables the model to learn the relative status of arbitrary interpolations for precise state positioning throughout the continuous transition. 
Second, we enforce semantic constraints on both the source-side and the target-side to anchor the trajectory endpoints. This dual-end alignment strategy serves to stabilize the latent manifold, preserving the structural integrity of user preferences throughout the training process

\subsubsection{Dual-Time Flow Modeling}
To adapt continuous flow modeling to the discrete item recommendation, we first project the discrete target item $i_{l+1}$ into a continuous manifold to serve as the generation endpoint. 
Specifically, we define the target embedding $\boldsymbol{x}_1$ as:
\begin{equation}
\boldsymbol{x}_1 \triangleq \boldsymbol{e}_{l+1} = \text{Embedding}(i_{l+1}),
\end{equation}
where $\text{Embedding}(\cdot)$ denotes a trainable embedding layer. 
Here, we set this embedding as $\boldsymbol{x}_1$, representing the target state towards which the flow evolves.

Next, the flow trajectory is constructed through a linear interpolation scheme defined in Eq.~\eqref{eq:define_inter}. 
Specifically, the initial state $\boldsymbol{x}_0$ is sampled from a Gaussian distribution $\mathcal{N}(0, I)$, and interpolated states $\boldsymbol{x}_t$ are generated by linear interpolating between $\boldsymbol{x}_0$ and $\boldsymbol{x}_1$.
During training, the model learns from these interpolated states along the trajectory, which encourages it to capture fine-grained variations in instantaneous preference transitions rather than overly smooth global trends.
To emphasize informative regions of the trajectory, we adopt the heavy-tailed sampling strategy from \citet{14} to sample the interpolation time $t$, which prioritizes high-frequency regions along the flow trajectory.

However, for these sampled states, the single time variable employed by existing flow-based methods is insufficient to distinguish velocity directions. 
This is because different user trajectories frequently share similar interpolated states yet diverge toward markedly different targets.
Addressing this limitation, we introduce a \textbf{dual-time parameterization} $(t, r)$, where $t$ encodes the interpolated state along the trajectory, and an additional fixed timestep $r = 1$ serves as a temporal anchor for the trajectory endpoint. 
This design enables the network to refine the instantaneous velocity by jointly accounting for the interpolated states and the remaining distance. 
We implement this by constructing a hybrid feature $\boldsymbol{\tau}$:
\begin{equation}
\boldsymbol{\tau} = \text{TimeEmb}(t) + \text{TimeEmb}(r - t),
\end{equation}
where $\text{TimeEmb}(\cdot)$ is a two-layer MLP. Note that while $r$ is fixed to 1 in Stage 1 to build the basic manifold, it becomes a critical variable in Stage 2 to facilitate trajectory linearity across arbitrary sub-intervals.

While this dual-time mechanism offers precise temporal anchors, relying solely on such deterministic cues can lead the model to memorize fixed trajectories rather than capturing the evolution of user interests.
To prevent overfitting, a stochastic input modulation strategy is adopted~\citet{9}. 
A noise vector $\boldsymbol{\lambda}$ is sampled layer-wise from $\mathcal{N}(\delta, \delta)$ to randomly scale the combined embeddings, where $\delta$ serves as a unified hyperparameter governing both the mean and variance of the perturbation.

Finally, by aggregating these preceding features, we formulate the total network input $\boldsymbol{E}$ as follows:
\begin{equation}
\boldsymbol{E} = \boldsymbol{e}_{\text{s}} + \boldsymbol{\lambda} \odot (\boldsymbol{x}_t + \boldsymbol{\tau}),
\end{equation}
where $\boldsymbol{e}_{\text{s}}$ denotes the interaction sequence embedding and $\odot$ represents element-wise multiplication.

With the complete flow trajectory and network input defined, the remaining challenge lies in learning a stable and informative velocity field.
Directly regressing the instantaneous velocity can be unstable due to its high variance across different interpolated states and target preferences.
To address this issue, drawing inspiration from the direct reconstruction strategy in diffusion models~\cite{80, 99}, the output of the flow model is reparameterized for improved optimization stability.
Instead of predicting the velocity vector explicitly, the network $f_{\theta}$ is designed to estimate the target state $\hat{\boldsymbol{x}}_1$.
Accordingly, the instantaneous velocity field $v_{\theta}$ is defined as:
\begin{equation}
v_{\theta}(\boldsymbol{x}_t, t, r) \triangleq f_{\theta}(\boldsymbol{x}_t, t, r) - \boldsymbol{x}_0.
\end{equation}

By substituting this formulation into the CFM objective, the dependency on the prior distribution $\boldsymbol{x}_0$ is eliminated. 
We refer to the resulting objective as the recovery loss, denoted by $\mathcal{L}_{rec}$:
\begin{equation}
\begin{aligned}
\mathcal{L}_{rec}(\theta)
&= \mathbb{E}_{t, \boldsymbol{x}_0, \boldsymbol{x}_1} \left\| (f_{\theta}(\boldsymbol{x}_t, t, 1) - \boldsymbol{x}_0) - (\boldsymbol{x}_1 - \boldsymbol{x}_0) \right\|^2  \\
&= \mathbb{E}_{t, \boldsymbol{x}_0, \boldsymbol{x}_1} \left\| f_{\theta}(\boldsymbol{x}_t, t, 1) - \boldsymbol{x}_1 \right\|^2.
\end{aligned}
\end{equation}
This derivation transforms the objective into a direct denoising regression problem. By explicitly anchoring the generation to the target item $\boldsymbol{x}_1$, this formulation avoids the complexity of modeling arbitrary distributions, ensuring the trajectory captures the evolution from uncertainty to concrete user intent.

\subsubsection{Dual-End Semantic Alignment}
However, relying solely on the recovery loss $\mathcal{L}_{rec}$ is insufficient for recommendation tasks. The learnable nature of item embeddings introduces a risk of representation collapse, where distinct embeddings merge to minimize reconstruction distance, thereby undermining item distinctiveness.

To address this issue, we enforce semantic alignment at both ends of the flow trajectory, constraining the model from complementary target-side and source-side perspectives. 
This dual-end design prevents representation collapse while stabilizing the latent manifold learned by the flow model.
At the target side, which corresponds to the embedding representation of the target item, we introduce a cross-entropy loss $\mathcal{L}_{tgt}$. This objective explicitly differentiates between various item embeddings, thereby preventing such representation collapse:
\begin{equation}
\begin{aligned}
\hat{y}_{l+1} &= \frac{\exp(f_{\theta}(\boldsymbol{x}_t, t, r) \cdot \boldsymbol{e}_{l+1})}{\sum_{j \in \mathcal{I}} \exp(f_{\theta}(\boldsymbol{x}_t, t, r) \cdot \boldsymbol{e}_j)}, \\
\mathcal{L}_{tgt} &= - \log \hat{y}_{l+1},
\end{aligned}
\end{equation}
where $\boldsymbol{e}_j$ denotes the embedding of item $j$, and $\hat{y}_{l+1}$ represents the normalized scores of predicting the target item $i_{l+1}$.

At the source side, which focuses on preserving the semantics of the user's historical interactions, we introduce a history reconstruction objective $\mathcal{L}_{src}$. To ensure the trajectory retains fidelity to the user's past behaviors, we reconstruct the observed interaction history $\mathbf{A}_u$ directly from the internal hidden state $\boldsymbol{E}_n$. This objective stabilizes the source semantics and anchors the trajectory's foundation:
\begin{equation}
\mathcal{L}_{src} = \left\| \mathcal{D}(\boldsymbol{E}_n) - \mathbf{A}_u \right\|^2,
\end{equation}
where $\boldsymbol{E}_n$ denotes the intermediate hidden representation and $\mathcal{D}(\cdot)$ denotes a decoding function that maps the predicted representation back to the raw interaction space. As a result, the source-side constraint stabilizes the starting semantics of the flow trajectory, providing a reliable reference point for subsequent generation.

By consistently enforcing this dual-end alignment throughout the training process, we explicitly anchor the semantic space to establish a robust manifold. This structural stability serves as a foundation for the subsequent single-step consolidation, ensuring that future efficiency gains do not come at the cost of representation quality.

\subsection{Training Stage 2: Single-Step Consolidation}
To resolve the trade-off between inference efficiency and recommendation accuracy, this stage consolidates the iterative flow trajectory into a direct generation mechanism. 
Specifically, we implement this consolidation with three strategies:
First, we introduce a semantic anchor prior for trajectory initialization. By initializing the flow with perturbed item embeddings rather than uninformative noise, we effectively shorten the semantic trajectory distance required for generation.
Second, we employ average velocity modeling for trajectory consolidation. We recast the flow objective to learn the global displacement vector, enabling the model to predict the target user intent in a single forward pass.
Finally, we enforce flow consistency regularization to maintain semantic stability by explicitly minimizing flow acceleration.

\subsubsection{Informative Prior Construction}
Building on the basic manifold constructed in the first stage, we explicitly freeze the embedding layer to preserve the established high-quality semantic space. 
Flow matching typically initiates generation from uninformative noise. However, this ignores the tight connection between a user's past behaviors and their future intentions, resulting in unnecessarily prolonged generation trajectories.
To overcome this, we substitute the noise distribution with an informative semantic anchor prior. Grounded in the insight that user interests evolve continuously, we initialize the flow using a semantic representation of past behaviors rather than random noise. This shortens the trajectory distance by transforming the generative task from creation from scratch to evolution from context.

Specifically, to prevent the model from learning trivial identity mappings from history to history, we perturb the historical item embedding with a Bernoulli-distributed binary mask. This perturbation compels the flow to reconstruct and evolve precise preference representations from partial behavioral cues, ensuring the model learns robust interest transitions. Therefore, we formulate the starting point $\boldsymbol{x}_0$ as follows:
\begin{equation}
\boldsymbol{x}_0 = \mathbf{m} \odot \boldsymbol{e}_{k}, \quad \text{where } k \sim \text{Uniform}(s_u), \ \mathbf{m} \sim \text{Bernoulli}(\rho),
\end{equation}
where $\boldsymbol{e}_{k}$ denotes the embedding of an item $k$ uniformly sampled from the interaction sequence $s_u$, and $\mathbf{m}$ represents a binary mask vector governed by a retention rate $\rho$.

\subsubsection{Average Velocity Field Establishment}
With the trajectory start point $\boldsymbol{x}_0$ optimized via the semantic anchor prior, the remaining challenge lies in efficiently traversing the manifold to reach the target preference $\boldsymbol{x}_1$.
In generative frameworks, this evolution is modeled by ODEs, where numerical solvers (e.g., the Euler method) approximate the trajectory through iterative updates:
\begin{equation}
\label{eq:ODE-solver}
\boldsymbol{x}_{t + \Delta t} = \boldsymbol{x}_t + v(\boldsymbol{x}_t, t) \cdot \Delta t.
\end{equation}
This forces an unavoidable trade-off, as the limit of $\Delta t \to 0$ necessary for precise preference modeling incurs prohibitive inference latency. 
Rather than focusing on accelerating numerical solvers, which fundamentally remain constrained by $\Delta t$, we revisit the flow formulation itself from a trajectory-level perspective. Specifically, we reformulate the flow relationship by directly integrating the ODE over the interval $[t, r]$:
\begin{equation}
\boldsymbol{x}_r - \boldsymbol{x}_t
= \int_{t}^{r} v_{\theta}(\boldsymbol{x}_{\eta}, \eta)\, d\eta
= (r - t)\times \left[\frac{1}{r - t}\int_{t}^{r} v_{\theta}(\boldsymbol{x}_{\eta}, \eta)\, d\eta \right],
\end{equation}
where $\eta$ denotes the continuous temporal variable corresponding to interpolated states along the trajectory. Observing that the dependence on the discretization step $\Delta t$ is eliminated, we identify the term in brackets as the average velocity $\bar{v}_{\theta}$ along the trajectory:
\begin{equation}
\bar{v}_{\theta}(\boldsymbol{x}_t, t, r) \triangleq \frac{1}{r - t}\int_{t}^{r} v_{\theta}(\boldsymbol{x}_{\eta}, \eta)\, d\eta.
\end{equation}

Consequently, Eq.~\eqref{eq:ODE-solver} can be reformulated as an exact algebraic relation, rather than an approximate differential description. Under the default flow where the trajectory starts at $t=0$ and ends at $r=1$, we obtain:
\begin{equation}
\boldsymbol{x}_1 = \boldsymbol{x}_0 + \bar{v}_{\theta}(\boldsymbol{x}_0, 0, 1).
\end{equation}
In this context, the average velocity acts as a global displacement that abstracts away redundant intermediate updates, enabling the model to consolidate the multi-step trajectory into a single transition that drives the informative prior toward the target item preference.

To realize this paradigm shift from instantaneous to average velocity modeling, we reformulate the training objective. 
By substituting our assumed average velocity $\bar{v}_{\theta}$ into the original flow matching constraint, the objective can be expanded as:
\begin{equation} 
\begin{aligned} 
\mathcal{L}_{CFM}(\theta) 
&= \mathbb{E}_{t, \boldsymbol{x}_0, \boldsymbol{x}_1} \left\| v_\theta(\boldsymbol{x}_t, t) - u_t(\boldsymbol{x}_t|\boldsymbol{x}_0, \boldsymbol{x}_1) \right\|^2 \\ 
&= \mathbb{E}_{t, \boldsymbol{x}_0, \boldsymbol{x}_1} \Bigg\| \bar{v}_{\theta}(\boldsymbol{x}_t, t, r) - (r-t)\frac{d}{dt}\bar{v}_{\theta}(\boldsymbol{x}_t, t, r)- (\boldsymbol{x}_1 - \boldsymbol{x}_0) \Bigg\|^2 \\ 
&= \mathbb{E}_{t, \boldsymbol{x}_0, \boldsymbol{x}_1} \Bigg\| \underbrace{ \bar{v}_{\theta}(\boldsymbol{x}_t, t, r) - (\boldsymbol{x}_1 - \boldsymbol{x}_0)}_{conditional~flow~matching} - \\
&\quad (r - t) \underbrace{\left[\frac{d\boldsymbol{x}_t}{dt} \cdot \frac{\partial}{\partial \boldsymbol{x}_t}\bar{v}_{\theta}(\boldsymbol{x}_t, t, r) + \frac{\partial}{\partial t}\bar{v}_{\theta}(\boldsymbol{x}_t, t, r)\right]}_{curvature~penalty} \Bigg\|^2.
\end{aligned} 
\end{equation}
Analyzing this objective reveals that the total error comprises two coupled terms: a conditional flow matching and a curvature penalty. 
Leveraging the triangle inequality, we can minimize the upper bound of this total error by optimizing its constituent norms separately, effectively decouple the training objective into two independent but cooperative tasks: ensuring precise item matching accuracy via the flow matching term, and guaranteeing extremely low inference latency via the curvature penalty.

\subsubsection{Flow Consistency Regularization}
First, we focus on the trajectory matching constraint. Departing from the fixed horizon $r \equiv 1$ of the initial stage, we compel the model to generalize across variable sub-intervals $[t, r]$ to explicitly enforce trajectory linearity, thereby preventing intermediate semantic deviations that would directly degrade recommendation accuracy in the absence of iterative refinement.
Specifically, to construct these intervals, we sample the start time $t \sim \text{Uniform}(0, 1)$ and determine the end time $r$ via a probabilistic mixture:
\begin{equation}
\label{eq:sampling_strategy}
r \sim p_{\text{end}} \cdot \delta_1 + (1 - p_{\text{end}}) \cdot \text{Uniform}(t, 1),
\end{equation}
where $p_{\text{end}}$ controls the probability of anchoring the end time to the trajectory terminus, and $\delta_1$ denotes a Dirac delta distribution centered at $1$.
Formally, we instantiate this requirement via a matching loss:
\begin{equation}
\label{eq:loss_match}
\mathcal{L}_{match} = \| f_{\theta}(\boldsymbol{x}_t, t, r) - \boldsymbol{x}_1 \|^2,
\end{equation}
this anchors the model's prediction directly to the target user interest, ensuring the global directional consistency required for one-step generation.

However, mere endpoint alignment does not strictly preclude interpolated distortions. While the matching loss ensures the destination is correct, it allows for non-linear fluctuations between temporal states.
Semantically, such fluctuations manifest as high curvature, implying unnecessary deviations through latent states unrelated to true preference evolution. This causes the trajectory to detach from the user preference manifold and leads to interest hallucinations.

To address this semantic deviation, we explicitly enforce a curvature penalty that minimizes flow acceleration $\frac{d}{dt}\bar{v}_{\theta}$.
By minimizing this acceleration, we explicitly eliminate the noise responsible for such deviations, ensuring precise one-step generation.

Crucially, under the employed reparameterization, since the starting state $\boldsymbol{x}_0$ is time-invariant, minimizing the flow acceleration $\frac{d}{dt}\bar{v}_{\theta}$ is mathematically equivalent to minimizing the total time derivative of the model's target prediction $\frac{d}{dt}f_{\theta}$.
We apply the chain rule to expand this total derivative with respect to the input variables $\boldsymbol{x}_t, t, \text{and } r$:
\begin{equation}
\frac{d}{dt} f_{\theta} = \frac{\partial f_{\theta}}{\partial \boldsymbol{x}_t} \frac{d\boldsymbol{x}_t}{dt} + \frac{\partial f_{\theta}}{\partial t} \frac{dt}{dt} + \frac{\partial f_{\theta}}{\partial r} \frac{dr}{dt}.
\end{equation}

By substituting the decomposed components ($\frac{dt}{dt} = 1$, $\frac{dr}{dt} = 0$), we recognize that the chain rule expansion is mathematically equivalent to the projection of the model's full Jacobian matrix along the trajectory tangent vector. However, explicitly instantiating the high-dimensional Jacobian is computationally prohibitive. Therefore, we employ the Jacobian-Vector Product (JVP) operator to efficiently compute this directional derivative in a single forward pass without materializing the matrix. Formally, the consistency loss is defined as:
\begin{equation}
\label{eq:loss_straight}
\mathcal{L}_{cons} = \Bigg\| \underbrace{[\bar{v}_{\theta}, 1, 0]}_{\text{Tangent Vector}} \cdot \underbrace{\nabla f_{\theta}(\boldsymbol{x}_t, t, r)}_{\text{Jacobian Matrix}} \Bigg\|^2,
\end{equation}
where $\nabla f_{\theta}$ denotes the full Jacobian. This strategy consolidates the evolution of user preferences, straightening the trajectory by minimizing flow acceleration to guarantee robust one-step inference.

Finally, the loss function utilized during this stage is formulated by integrating four distinct components, represented as follows:
\begin{equation}
    \mathcal{L} = \mathcal{L}_{match} + \gamma \mathcal{L}_{cons} + \alpha \mathcal{L}_{tgt} + \beta \mathcal{L}_{src},
\end{equation}
where $\alpha$, $\beta$, and $\gamma$ are the hyperparameters.

\subsection{Model Inference}
During the inference phase, we initialize the starting state $\boldsymbol{x}_0$ by directly applying the masking mechanism to the user's historical interaction embedding, strictly mirroring the protocol used in the training stage 2.
Building upon this initialization, the strict linear trajectory characteristics enforced by consistency optimization facilitate a rapid one-step generation mechanism. Specifically, by setting the time step $\Delta t = 1$, the predicted embedding of the target item $\boldsymbol{\hat{x}}_1$ is resolved via a single forward pass:
\begin{equation}
\boldsymbol{\hat{x}}_1 = \boldsymbol{x}_0 + \bar{v}_{\theta}(\boldsymbol{x}_0, 0, 1) \equiv f_{\theta}(\boldsymbol{x}_0, 0, 1).
\end{equation}
Finally, relevance scores between $\boldsymbol{\hat{x}}_1$ and all candidate item embeddings are computed via inner product. The top-K items with the highest scores are recommended.

\section{Experimental Setup}
\subsection{Dataset \& Baselines}
To evaluate \model on diverse data settings, we conduct experiments on three benchmarks: ML-100k\footnote{\url{http://files.grouplens.org/datasets/movielens/}}, Amazon-Beauty\footnote{\url{https://cseweb.ucsd.edu/~jmcauley/datasets/amazon_v2/}}, and Steam\footnote{\url{https://steam.internet.byu.edu/}}.
These datasets vary in size and sparsity, ranging from the dense ML-100k to the large and sparse Amazon-Beauty and Steam datasets.
We use the same data pre-processing method as used in ICLRec~\cite{22} and DuoRec~\cite{23}.

We compare \model against three category baselines: \textbf{(1) RNN/CNN-based models:} GRU4Rec~\cite{2} utilizes GRU to capture in-session behavioral patterns and predict subsequent user item preferences, and Caser~\cite{3} leverages convolutional neural networks to map user action sequences into both temporal and latent spaces; 
\textbf{(2) Transformer-based models:} SASRec~\cite{4} introduces a self-attention based decoder architecture to effectively capture long-term dependencies, BERT4Rec~\cite{5} employs a bidirectional transformer architecture coupled with a Cloze task to learn users’ dynamic preferences, and STOSA~\cite{6} utilizes Wasserstein attention to model uncertainty for accurate representation of evolving user preferences; and 
\textbf{(3) Generative models:} AutoSeqRec~\cite{7} leverages a multi-autoencoder framework to fuse long-term user preferences and short-term interests, DreamRec~\cite{8} achieves direct generation of personalized oracle item embeddings through a guided diffusion model, DiffuRec~\cite{9} models items as distributions using a diffusion model to capture multi-faceted content and user preferences, and FMRec~\cite{10} employs flow matching with straight trajectories to eliminate random noise in recommendation generation.


\begin{table*}[htbp]
  \centering
  \caption{Overall performance comparison. Subscripts indicate the performance gap relative to the baseline model in each category. We highlight the highest-performing metric values in \textbf{bold} and the second-best values in \underline{underlined}.}
  \label{tab:main}
  
  \setlength{\tabcolsep}{2.5pt} 
  \footnotesize 
  
  \newcommand{\redgap}[1]{$_{\!\textcolor{myred}{#1}}$}
  \newcommand{\greengap}[1]{$_{\!\textcolor{mygreen}{#1}}$}
  
  \newcommand{\nogap}{\phantom{$_{\!-0.00}$}}
  
  \newcommand{\bottomstrut}{\rule[-1.2ex]{0pt}{3.5ex}}

  \resizebox{\textwidth}{!}{
    \begin{tabular}{lcccccccccccc}
    \toprule
    \multirow{2}{*}{\textbf{Model}} & \multicolumn{4}{c}{\textbf{ML-100k}} & \multicolumn{4}{c}{\textbf{Beauty}} & \multicolumn{4}{c}{\textbf{Steam}} \\
    \cmidrule(lr){2-5} \cmidrule(lr){6-9} \cmidrule(lr){10-13}
          & H@10 & H@20 & N@10 & N@20 & H@10 & H@20 & N@10 & N@20 & H@10 & H@20 & N@10 & N@20 \\
    \midrule
    \multicolumn{13}{l}{\textit{\textbf{RNN/CNN-based Models}}} \\ 
    \midrule
    GRU4Rec & 12.1951\nogap & 21.8451\nogap & 6.0326\nogap & 8.4727\nogap & 1.9370\nogap & 3.8531\nogap & 0.9029\nogap & 1.3804\nogap & 5.4257\nogap & 9.2319\nogap & 2.6033\nogap & 3.5572\nogap \\
    Caser & 11.2426\redgap{-0.95} & 19.5189\redgap{-2.33} & 5.0683\redgap{-0.96} & 7.1439\redgap{-1.33} & 2.8166\greengap{+0.88} & 4.4048\greengap{+0.55} & 1.3602\greengap{+0.46} & 1.7595\greengap{+0.38} & 6.4940\greengap{+1.07} & 10.9633\greengap{+1.73} & 3.0846\greengap{+0.48} & 4.2043\greengap{+0.65} \\
    \midrule
    \multicolumn{13}{l}{\textit{\textbf{Transformer-based Models}}}  \\ 
    \midrule
    SASRec & 13.5737\nogap & 22.6935\nogap & 6.3427\nogap & 8.6340\nogap & 6.2648\nogap & 8.9791\nogap & 3.2305\nogap & 3.6563\nogap & 8.3763\nogap & 13.6060\nogap & 4.0489\nogap & 5.3630\nogap \\
    BERT4Rec & 9.3319\redgap{-4.24} & 16.8611\redgap{-5.83} & 4.4568\redgap{-1.89} & 6.3442\redgap{-2.29} & 3.7160\redgap{-2.55} & 5.7922\redgap{-3.19} & 1.8291\redgap{-1.40} & 2.3541\redgap{-1.30} & 7.9448\redgap{-0.43} & 12.7322\redgap{-0.87} & 4.0002\redgap{-0.05} & 5.2027\redgap{-0.16} \\
    STOSA & 13.6542\greengap{+0.08} & 21.7761\redgap{-0.92} & 5.2159\redgap{-1.13} & 8.3302\redgap{-0.30} & 6.1262\redgap{-0.14} & 9.0954\greengap{+0.12} & 3.2053\redgap{-0.03} & 3.9491\greengap{+0.29} & 8.5870\greengap{+0.21} & 14.1107\greengap{+0.50} & 4.1191\greengap{+0.07} & 5.5072\greengap{+0.14} \\
    \midrule
    \multicolumn{13}{l}{\textit{\textbf{Generative-based Models}}}  \\ 
    \midrule
    DiffuRec & 12.8501\nogap & 19.4127\nogap & 6.4136\nogap & 8.0459\nogap & 7.8374\nogap & 10.9358\nogap & 4.6971\nogap & 5.4784\nogap & 9.8437\nogap & 15.3817\nogap & 5.0429\nogap & 6.4340\nogap \\
    AutoSeqRec & 14.6641\greengap{+1.81} & 22.8724\greengap{+3.46} & 7.3955\greengap{+0.98} & 9.4584\greengap{+1.41} & 7.1016\redgap{-0.74} & 9.3342\redgap{-1.60} & 4.0157\redgap{-0.68} & 5.0133\redgap{-0.47} & 8.7741\redgap{-1.07} & 14.6752\redgap{-0.71} & 4.4729\redgap{-0.57} & 5.9823\redgap{-0.45} \\
    DreamRec & 12.4377\redgap{-0.41} & 20.8357\greengap{+1.42} & 5.9837\redgap{-0.43} & 7.8234\redgap{-0.22} & 6.9821\redgap{-0.86} & 9.4531\redgap{-1.48} & 3.9769\redgap{-0.72} & 4.9860\redgap{-0.49} & 8.9875\redgap{-0.86} & 15.0871\redgap{-0.29} & 4.6416\redgap{-0.40} & 5.9701\redgap{-0.46} \\
    FMRec & \underline{15.4934}\greengap{+2.64} & \underline{24.3146}\greengap{+4.90} & \underline{7.6571}\greengap{+1.24} & \underline{9.8158}\greengap{+1.77} & \underline{8.2693}\greengap{+0.43} & \underline{11.626}\greengap{+0.69} & \underline{4.9461}\greengap{+0.25} & \underline{5.7876}\greengap{+0.31} & \underline{10.5908}\greengap{+0.75} & \underline{16.4669}\greengap{+1.09} & \underline{5.4925}\greengap{+0.45} & \underline{6.9689}\greengap{+0.53} \\
    \rowcolor{lightgray} 
    \textbf{\model} \bottomstrut & \textbf{15.8722}\greengap{+3.02} & \textbf{26.1340}\greengap{+6.72} & \textbf{8.1968}\greengap{+1.78} & \textbf{10.7879}\greengap{+2.74} & \textbf{8.4834}\greengap{+0.65} & \textbf{11.9048}\greengap{+0.97} & \textbf{4.9744}\greengap{+0.28} & \textbf{5.8351}\greengap{+0.36} & \textbf{10.7080}\greengap{+0.86} & \textbf{16.6562}\greengap{+1.27} & \textbf{5.4973}\greengap{+0.45} & \textbf{6.9911}\greengap{+0.56} \\
    \bottomrule
    \end{tabular}%
  }
\end{table*}

\subsection{Metrics \& Implementation Details}

We employ two widely used metrics: Hit Rate (H@K) and Normalized Discounted Cumulative Gain (N@K). 
Following the established all-ranking protocol~\cite{11}, we rank all non-interacted items for each user and report results for K$ \in \{10, 20\}$.

We implement the backbone flow model using a transformer-based architecture with 4 self-attention heads and a hidden dimension of $128$. 
The decoder is a 3-layer MLP with the $\tanh$ activation function. 
We set the batch size to $512$ and the maximum user interaction sequence length to $50$.

\begin{table}[t]
    \centering
    \caption{Ablation study.}
    \label{tab:ablation}
    \small
    
    \begin{tabular*}{\columnwidth}{@{\extracolsep{\fill}}lcccc}
        \toprule
        \textbf{Datasets} & \multicolumn{2}{c}{\textbf{ML-100k}} & \multicolumn{2}{c}{\textbf{Beauty}} \\
        \cmidrule(r){1-5}
        \textbf{Metrics} & H@20 & N@20 & H@20 & N@20 \\
        \midrule
        $\text{w/o~stage}$ & 25.2610 & 10.4645 & 11.7177 & 5.7838 \\
        $\text{w/o~} \mathcal{L}_{cons}$   & 24.8579 & 10.3656 & 11.8406 & 5.8031 \\
        $\text{w/o~} \mathcal{L}_{match}$         & 21.7264 & 8.8802  & 11.7805 & 5.8086 \\
        $\text{w/o~prior}$  & 20.6521 & 8.8471  & 11.7939 & 5.7417 \\
        \midrule
        \textbf{\model} & \textbf{26.1340} & \textbf{10.7879} & \textbf{11.9048} & \textbf{5.8351} \\
        \bottomrule
    \end{tabular*}
\end{table}

\section{Experimental Results}
\subsection{Overall Performance}
Table~\ref{tab:main} presents the overall performance comparison of \model and the baselines across three datasets. 
The generative sequential models demonstrate better performance compared to non-generative baseline methods such as SASRec and BERT4Rec. 
This suggests that generative SR models are promising for capturing user preferences, as they can effectively model the complex evolution of user-item interactions through a generation paradigm.
We observe that flow-matching approaches (\eg FMRec and \model), consistently outperform matrix-reconstruction models such as AutoSeqRec and stochastic diffusion methods like DiffuRec. 
In contrast to the limitations of discrete constraints and stochastic perturbations inherent in these baselines, flow-based methods model preference evolution as a deterministic ODE trajectory in continuous space. 
This paradigm effectively rectifies the transport trajectory, providing a more stable and expressive representation.

Our proposed \model consistently outperforms all baseline models across all datasets and metrics. Notably, on the ML-100k dataset, \model surpasses the strongest state-of-the-art competitor FMRec by a significant margin (with p-value $<$ 0.01), achieving a relative improvement of 9.90\% in N@20 and 7.48\% in H@20. 
This demonstrates that \model effectively addresses the limitations of flow matching. 
By incorporating the semantic anchor prior to shorten the generation trajectory and modeling the global average velocity to eliminate linear redundancy, \model constructs a more direct and accurate mapping, leading to more precise and efficient personalized recommendations.

\subsection{Ablation Study}
To verify the effectiveness of each component in \model, we conduct an ablation study on the ML-100k and Beauty datasets. 
Specifically, the impact of key components is evaluated through four distinct variants: (1) the variant without the second training stage is labeled as $\text{w/o~stage}$; (2) the variant excluding the trajectory consistency loss is labeled as $\text{w/o~}\mathcal{L}_{cons}$; (3) the variant removing the matching consistency loss is labeled as $\text{w/o~}\mathcal{L}_{match}$; and (4) the variant replacing the semantic anchor prior with Gaussian noise is labeled as $\text{w/o~prior}$. 
The experimental results are presented in Table~\ref{tab:ablation}.

From the results, we can observe that: 
(1) The matching consistency loss $\mathcal{L}_{match}$ and the semantic anchor prior represent the most fundamental module of \model. Their removal leads to the most pronounced performance degradation. This confirms that the core flow matching objective and a semantically informed starting point are critical. Specifically, they enable the model to fully leverage existing semantic information to transport the initial distribution to the target distribution. 
(2) Replacing the semantic anchor prior with Gaussian noise ($\text{w/o~prior}$) results in a significant decline in effectiveness. This indicates that a curated semantic anchor provides a more efficient shortcut through the latent space than random noise, simplifying the generative trajectory and reducing the learning difficulty. 
(3) Removing either the trajectory consistency loss $\mathcal{L}_{cons}$ or the second training stage results in a certain degree of performance degradation. This decline arises because both components primarily regularize the dynamics of user preference evolution within an established latent item space. They do not attempt to reconstruct the fundamental semantic representations of items. Therefore, the performance decline is less pronounced compared to the sharp degradation observed with $\text{w/o~} \mathcal{L}_{match}$ or  ($\text{w/o~prior}$).
(4) Ablation effects are more evident on denser datasets, while performance variations on sparse datasets remain limited. In extremely sparse scenarios, weak interaction signals constrain the benefits of fine-grained trajectory constraints, making core components more influential than advanced refinements.
These findings underscore the necessity of both foundational mapping and trajectory-level consistency for robust one-step generative recommendation.

\begin{table}[!t]
  \centering
  \caption{Inference efficiency comparison. Metrics are reported per sample for FLOPs and Latency.}
  \label{tab:efficiency}
  \small
  
  \begin{tabular*}{\columnwidth}{@{\extracolsep{\fill}}clccc}
    \toprule
    \textbf{Dataset} & \textbf{Metric} & \textbf{DiffuRec} & \textbf{FMRec} & \textbf{\model} \\
    \midrule
    \multirow{3}{*}{\textbf{ML-100k}} 
          & FLOPs (G) & 1.26 & 1.32 & \textbf{0.04} \\
          & Latency (ms) & 105.97 & 98.04 & \textbf{4.44} \\
          & Infer. Time (s) & 2.57 & 2.34 & \textbf{0.10} \\
    \midrule
    \multirow{3}{*}{\textbf{Beauty}} 
          & FLOPs (G) & 1.26 & 1.96 & \textbf{0.07} \\
          & Latency (ms) & 103.19 & 93.74 & \textbf{4.49} \\
          & Infer. Time (s) & 42.21 & 40.44 & \textbf{2.25} \\
    \midrule
    \multirow{3}{*}{\textbf{Steam}} 
          & FLOPs (G) & 1.26 & 2.02 & \textbf{0.07} \\
          & Latency (ms) & 104.45 & 92.81 & \textbf{3.99} \\ 
          & Infer. Time (s) & 507.23 & 516.27 & \textbf{26.23} \\
    \bottomrule
  \end{tabular*}
\end{table}

\begin{figure}[!t]
    \centering
    \includegraphics[width=1\linewidth]{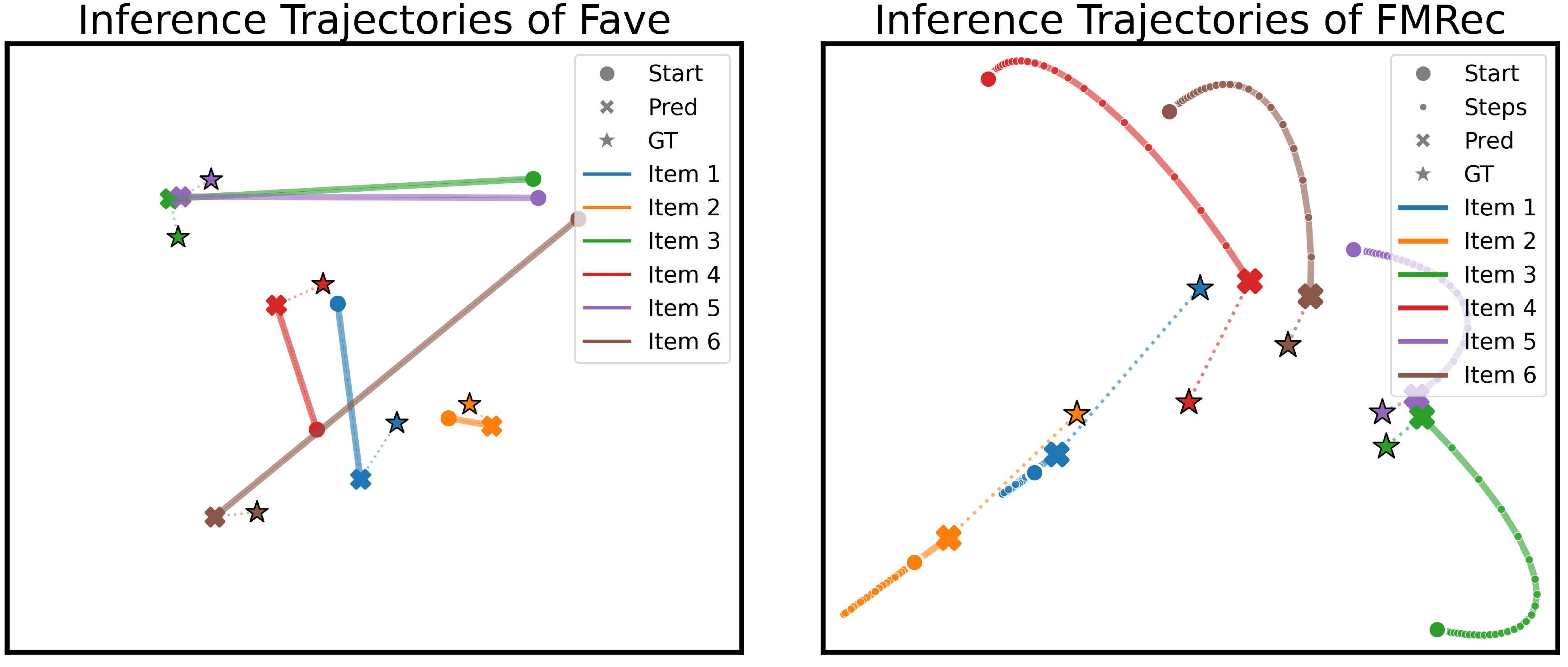}
    \caption{Comparison of inference trajectories for \model (left) and FMRec (right) on ML-100k. The trajectory points are first $\ell_2$-normalized and then visualized using t-SNE.}
    \label{fig:trajectory}
\end{figure}

\subsection{Efficiency Analysis}
To evaluate the efficiency of \model, we conduct experiments on three datasets with varying interaction scales: ML-100k, Beauty, and Steam. Specifically, the evaluation benchmarks our model against the diffusion-based DiffuRec and the flow-based FMRec in terms of FLOPs, Latency, and Inference Time. All experiments are conducted on a single NVIDIA L40 GPU under a strictly controlled baseline setting to ensure fairness.

As shown in Table~\ref{tab:efficiency}, generative baselines like DiffuRec and FMRec suffer from high computational overhead due to their reliance on iterative numerical solvers. Although FMRec improves trajectory straightness, it still necessitates multiple steps to traverse the full generation trajectory from noise to data. In contrast, \model achieves a major efficiency breakthrough, delivering a speedup of over 20 times on the ML-100k dataset relative to state-of-the-art generative baselines. This efficiency stems from our one-step generation mechanism, which utilizes the semantic anchor prior to bypass iterative integration, enabling direct preference prediction in a single forward pass. Thus, \model effectively resolves the latency bottleneck, proving highly practical for real-time recommendation.

\subsection{Trajectories Visualization Analysis}
To analyze the evolution of preference representations during inference, we sample several target items from the ML-100k test set, first $\ell_2$-normalize the trajectory points, and then visualize them using t-SNE to compare \model with FMRec.
In Figure~\ref{fig:trajectory}, FMRec is evaluated under its optimal setting of a 30-step iterative process. Conversely, \model completes inference in a single step. 
Despite this reduction in inference steps, \model consistently produces predictions that lie very close to the ground-truth targets. 
In contrast, FMRec exhibits notably less stable behavior during the inference process. 
For items 1, 2, and 3, the trajectories contain points with substantial deviations from the target items, even within the full 30-step trajectory. 
Moreover, certain generated trajectories terminate in close proximity to their starting points. 
This implies a failure to evolve the initial noise into distinct user preferences, leaving the final representations uninformative.
Overall, the visualization demonstrates that \model achieves more direct, stable, and accurate preference transitions while using a single generation step, highlighting our superior inference efficiency and robustness compared to FMRec.

\begin{figure}[!t]
    \centering
    \includegraphics[width=0.8\linewidth]{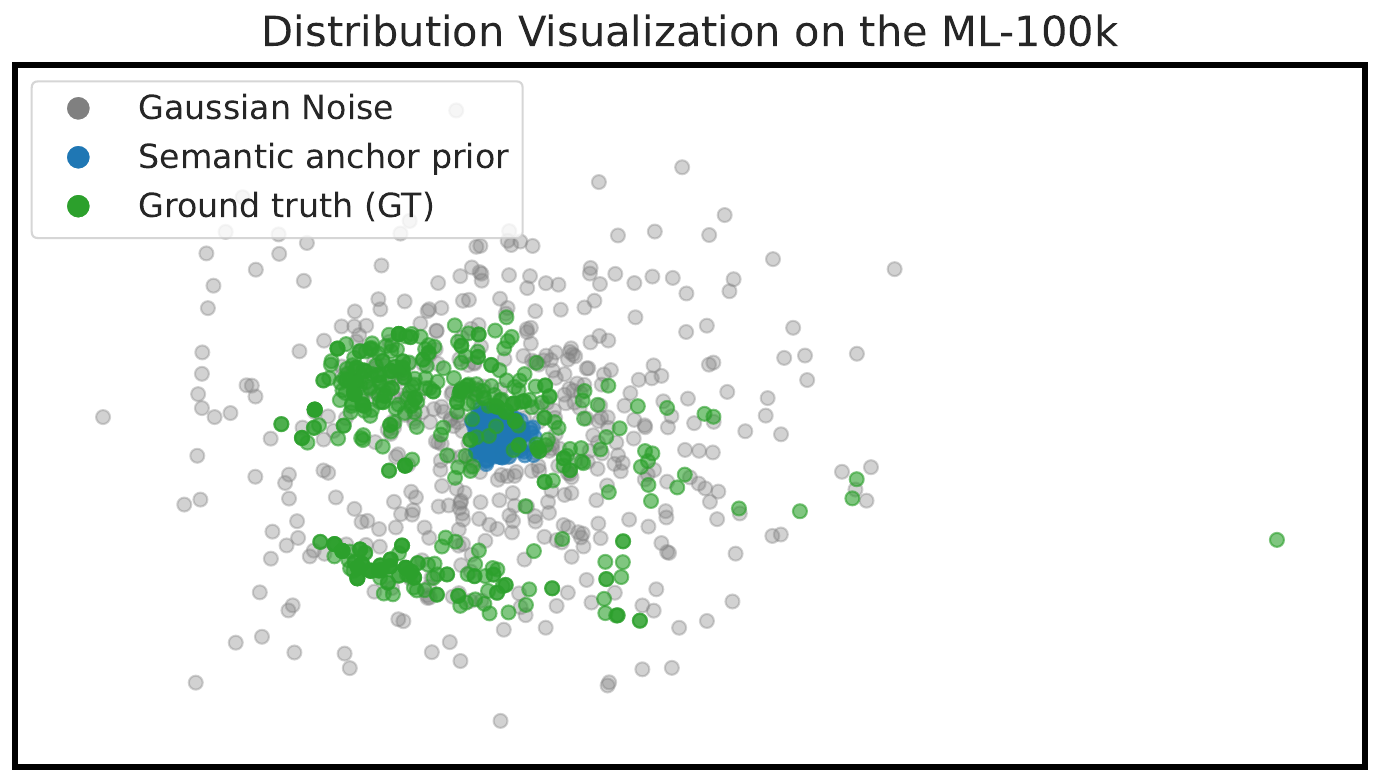}
    \caption{Visualization of embedding distributions on ML-100k for Gaussian noise, the proposed semantic anchor prior, and ground-truth item embeddings, showing that the semantic anchor prior is initialized closer to the target distribution.}
    \label{fig:distribution}
\end{figure}

\subsection{Embedding Distribution Visualization}

To investigate whether the proposed semantic anchor prior provides a more suitable initialization for recommendation generation, we visualize and compare the embedded distributions of Gaussian noise, our semantic anchor prior, and the ground truth (GT). 
As shown in Figure~\ref{fig:distribution}, Gaussian noise is broadly distributed across the latent space without clear structure, whereas the GT embeddings form two compact clusters.
Within each cluster, items correspond to similar user demands or consumption intents. 
The semantic anchor prior overlaps with one of the ground-truth clusters in the embedding space. This alignment suggests that the semantic anchor prior captures meaningful user demands and provides a more consistent, lower-variance initialization.
In contrast, Gaussian noise is highly dispersed and sensitive to random fluctuations. 
As a result, the semantic anchor offers a more reliable starting point for trajectory generation, which shortens the trajectory distance. 
This supports more stable and effective recommendation modeling.

\begin{table}[!t]
  \centering
  \caption{Recommendation diversity performance on the ML-100k dataset.}
  \label{tab:diversity}
  \small 
  \renewcommand{\arraystretch}{1.1}
  
  \begin{tabular*}{\columnwidth}{@{\extracolsep{\fill}}llcc}
    \toprule
    \textbf{Method} & \textbf{Category} & \textbf{ILD} & \textbf{NDCG@20} \\
    \midrule
    Caser      & CNN-based         & 0.7440 & 7.1439 \\
    SASRec     & Transformer-based & 0.8866 & 8.6340 \\
    AutoSeqRec & AutoML-based      & 0.3350 & 9.4584 \\
    DiffuRec   & Diffusion-based   & 0.8313 & 8.0459 \\
    FMRec      & Flow-based        & 0.3786 & 9.8158 \\
    \midrule
    \textbf{\model} & \textbf{Flow-based} & 0.4348 & 10.7879 \\
    \bottomrule
  \end{tabular*}
\end{table}

\begin{figure}
    \centering
    \includegraphics[width=1\linewidth]{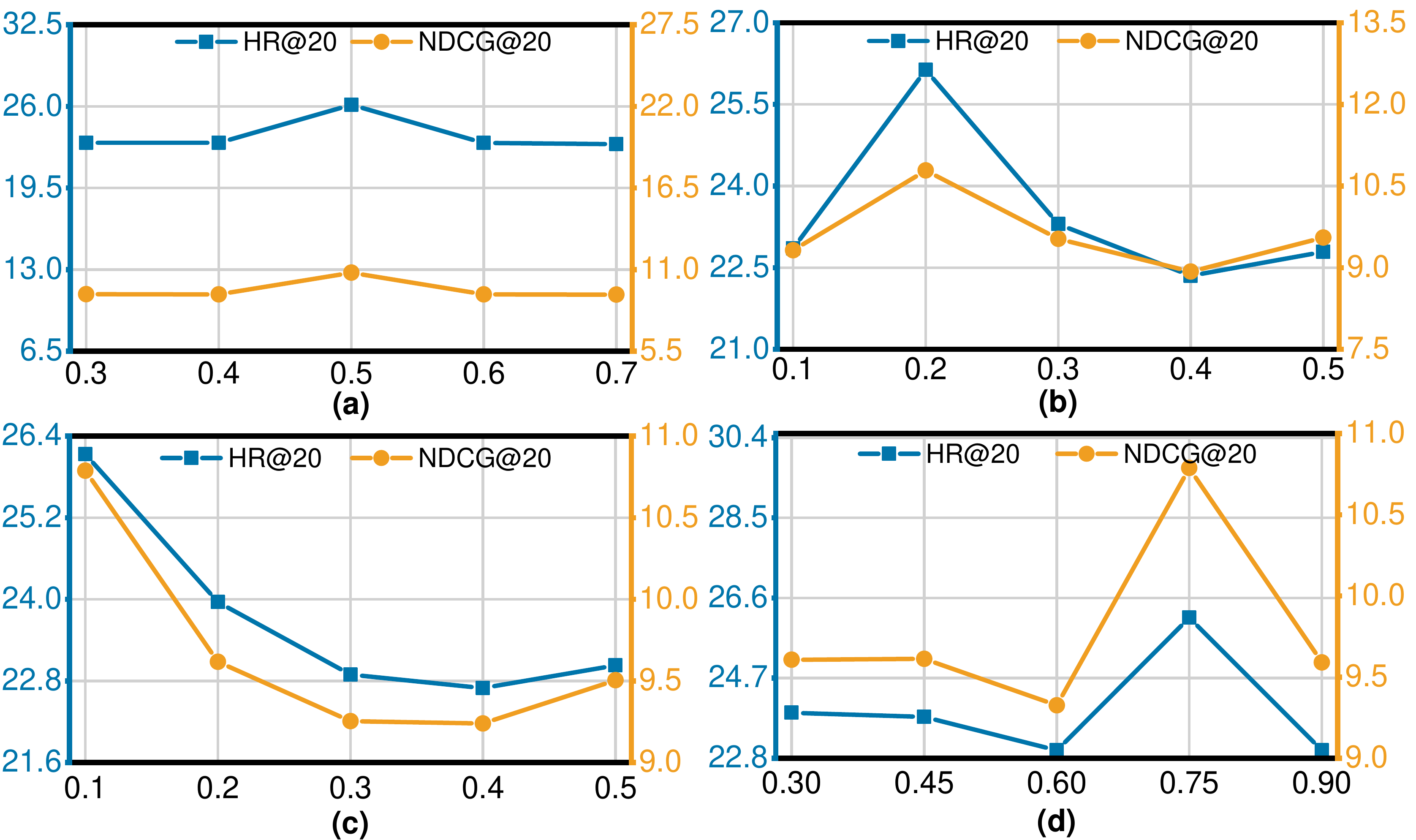}
    \caption{Performance of different hyperparameters. (a) weight $\alpha$ for $\mathcal{L}_{tgt}$, (b) weight $\beta$ for $\mathcal{L}_{src}$, (c) weight $\gamma$ for $\mathcal{L}_{cons}$, and (d) retention rate $\rho$.}
    \label{fig:hyperpameter}
\end{figure}

\subsection{Diversity Analysis}

To verify that our one-step paradigm retains generative diversity rather than collapsing into deterministic point estimation, we evaluated the Intra-List Diversity (ILD) alongside recommendation accuracy NDCG@20 on the ML-100k dataset. 
Specifically, ILD measures the average pairwise cosine distance between items within the top-20 list. 
The results are summarized in Table~\ref{tab:diversity}.

The diffusion-based DiffuRec yields high diversity but suffers from suboptimal accuracy, suggesting its diversity stems primarily from stochastic noise rather than precise preference modeling. 
A more critical comparison regarding mode collapse arises between our one-step \model and the multi-step FMRec. 
Although single-step generation theoretically risks degenerating into deterministic point estimation, our \model achieves notably higher diversity scores than both FMRec and AutoSeqRec. 
This confirms that our explicit modeling of average velocity effectively prevents distribution collapse. 
Unlike FMRec, which tends to yield over-smoothed trajectories to minimize transport cost, our approach utilizes the average velocity to enforce strict trajectory consistency. 
This ensures that for any arbitrary point sampled from the prior distribution, it is guided along a unique, rectified trajectory to transport onto a distinct region of the target distribution.  
Consequently, \model captures the full data distribution rather than collapsing into a point estimate, yielding both high diversity and accuracy.

\subsection{Analysis of Hyperparameters}

\textbf{Effectiveness of $\alpha$ weight of $\mathcal{L}_{tgt}$.} 
As illustrated in Figure~\ref{fig:hyperpameter}(a), the model achieves optimal performance when the target-aware cross-entropy loss $\mathcal{L}_{tgt}$ is weighted at $\alpha = 0.5$. When $\mathcal{L}_{tgt}$ is overemphasized, predicted item distributions across different users become highly similar, resulting in reduced recommendation diversity. In contrast, insufficient constraints on target items lead to unstable training and degraded recommendation quality. 

\textbf{Effectiveness of $\beta$ weight of $\mathcal{L}_{src}$.} 
As illustrated in Figure~\ref{fig:hyperpameter}(b), the history reconstruction loss $\mathcal{L}_{src}$ achieves its best performance at $\beta = 0.2$. When the weight is too small, the reconstruction constraint on the user interaction vector $\mathbf{A}_u$ is insufficient. The denoised representation then becomes more sensitive to noise and less effective at capturing user preferences, which degrades recommendation accuracy. In contrast, an excessively large weight overemphasizes historical interaction reconstruction and limits the model’s ability to adapt representations for next-item prediction. 

\textbf{Effectiveness of $\gamma$ weight of $\mathcal{L}_{cons}$.} 
As shown in Figure \ref{fig:hyperpameter}(c), performance peaks when the curvature regularization weight is set to $\gamma = 0.1$. When the penalty is too small, the generation trajectory deviates from linearity, causing inconsistent predictions across time steps and reducing the reliability of one-step recommendations. In contrast, an excessively large penalty over-constrains the model, limiting its capacity to adapt predictions to user-specific preferences. 

\textbf{Effectiveness of $\rho$~ retention rate of the semantic anchor mask.}
As shown in Figure \ref{fig:hyperpameter}(d), performance peaks when the retention rate is set to $\rho = 0.75$.
When the retention rate is too low, insufficient semantic information is retained in the initial prior distribution, leading to unstable generation trajectories and inconsistent one-step predictions.
In contrast, an excessively high retention rate preserves too much information, resulting in trivial identity mappings that weaken effective trajectory learning.

\section{Conclusion}
In this paper, we introduce \model, a novel one-step generative sequential recommendation framework. 
Our methodology is organized through a progressive two-stage training strategy. 
In Stage 1, we establish a robust latent preference space. In this stage, we propose dual-end semantic alignment which applies constraints at both the source (user history) and target (next item) ends of the flow trajectory. 
This stabilizes the learned manifold and prevents representation collapse. 
In Stage 2, we directly tackle the efficiency bottlenecks by introducing the semantic anchor prior, which replaces random noise with a masked embedding from the user's interaction history, providing an informative starting point close to the target. 
We then model the global average velocity, effectively consolidating the multi-step trajectory into a single displacement vector, which minimizes flow acceleration to ensure a consistent and direct generation trajectory.
Experiments on three benchmarks demonstrate that \model consistently achieves state-of-the-art recommendation performance while delivering an order-of-magnitude improvement in inference efficiency. 
This work bridges the gap between the high performance of generative models and the strict latency requirements of real-world recommender systems.

\section*{Acknowledgments}
This work was supported by the National Natural Science Foundation of China (62432002, 62406061, U25B2049), the National Key R\&D Program of China (2024YFE0111800), and the State Key Laboratory of Internet Architecture (HLW2025MS10).

\bibliographystyle{ACM-Reference-Format}
\bibliography{ref}
\end{document}